# Emergent Anomalous and Topological Hall Responses in an Epitaxial Ferromagnetic Weyl Nodal-Line metal $Fe_5Si_3$

Shubhashish Pati[1], Sonali Srotaswini Pradhan[2], Abhay Pandey[1], Nikita Sharma[1], Nanhe Kumar Gupta[a], Nakul Kumar[1], Nidhi Shukla[1], Saurav Singh[1], Vidhi Jain[1], Mitali[1], V. Kanchana[2,*], Sujeet Chaudhary[1,*]

[1] Thin Film Laboratory, Department of Physics, Indian Institute of Technology Delhi, New Delhi 110016, India

[2] Department of Physics, Indian Institute of Technology Hyderabad, Kandi 502285, Sangareddy, Telangana, India.

## Abstract

The interplay between real and reciprocal space topology yields intrinsically linked transport phenomena in magnetic Weyl systems, wherein the broken time-reversal symmetry, strong Dzyaloshinskii-Moriya interaction, and pronounced uniaxial anisotropy stabilize the momentum-space Berry-curvature monopoles (*Weyl nodes*) and real-space chiral spin textures. We present a combined first-principles and experimental study of epitaxial $Fe_5Si_3$ thin films, establishing them as a magnetic Weyl nodal-line material. First-principles Density Functional Theory (DFT) calculations unambiguously reveal that $Fe_5Si_3$ hosts a topologically nontrivial electronic structure containing six pairs of Weyl nodes at or near the Fermi level, accompanied by pronounced Berry curvature at high-symmetry points of the Brillouin Zone. High-quality epitaxial films exhibit robust ferromagnetism with a Curie temperature of ~370 K and strong magneto crystalline anisotropy. The magneto transport measurements on epitaxial films reveal the corresponding Berry curvature-driven responses, including a significantly large intrinsic anomalous Hall conductivity of 504 S/cm and a high anomalous Hall angle of 5.5%, which is in good agreement with DFT calculations. A negative and non-saturating longitudinal magnetoresistance is observed, consistent with a chiral-anomaly contribution from Weyl fermions near the Fermi level ($E_F$). Furthermore, a substantial topological Hall resistivity of 1.6 μΩ cm robust across a wide temperature range, indicating the possibility of robust chiral spin textures in the thin-film geometry. These combined theoretical and experimental results establish $Fe_5Si_3$ as a unique, low-cost, centrosymmetric magnetic Weyl nodal-line material, providing a versatile platform for exploring coupled real and reciprocal space topologies in topological spintronic applications.

Corresponding Authors

E-mail address: sujeetc@physics.iitd.ac.in (S. Chaudhary),

kanchana@phy.iith.ac.in (V. Kanchana)

[a] Present address: Centre for Magnetic and Spintronic Materials (CMSM), National Institute for Materials Science (NIMS) 1-2-1 Sengen, Tsukuba, Ibaraki 305-0047, Japan.

## Introduction

The intertwined emergence of real-space topology through skyrmions, antiskyrmions, and other chiral spin textures[1,2], and momentum-space topology encoded by Berry curvature arising from topologically nontrivial band crossings[3–5], unambiguously position *dual topology* as a defining framework for next-generation topological spintronics. This duality establishes a direct, symmetry-governed coupling between the emergent electromagnetic fields associated with chiral spin textures and the Berry curvature concentrated near band degeneracies, thereby *amplifying* the quantum-transport phenomena such as the Anomalous Hall Effect (AHE), whose magnitude is governed by the polarity of Berry-curvature, and the Topological Hall Effect (THE), which govern the fingerprint of the topology of chiral spin textures[6,7]. Recent material discoveries vividly demonstrate this synergy: the recently explored hexagonal Kagome magnet $Fe_3Ge$ exhibits strong real and reciprocal space Berry curvature effects, yielding an AHC of 550 S/cm and a sizable THE resistivity of 0.9 μΩ cm[8]. Magnetic Weyl nodal-ring semimetals such as $Mn_5Ge_3$ host multiple nodal rings near the $E_F$ in conjunction with Néel-type skyrmion lattices, producing an AHC of 980 S/cm in a particular crystallographic direction and THE amplitudes ~ 0.9 μΩ cm[6]. Likewise, $Mn_2Pd_{0.5}Ir_{0.5}Sn$ combines 42 Weyl-node pairs with room-temperature antiskyrmion phases, giving rise to a moderate AHC of 110 S/cm and exceptionally large THE resistivity up to 6 μΩ cm[7]. Two-dimensional van der Waals magnets, such as $Fe_3GeTe_2$, exhibit an additional degree of tunability, as thickness-dependent Berry-curvature modulation enables electric-field control over spin chiral structures[9][10].

Despite these advances, substantial challenges persist. Only a narrow class of materials exhibits robust dual topology under ambient conditions, and many require deliberate compositional tuning, such as Ir substitution in $Mn_2PdSn$ to stabilize their topological phases[7]. In several cases, the Curie temperatures of these systems also fall below room temperature[6,10], further limiting their technological applicability. Moreover, the majority of reported dual-topological materials have been realized exclusively in single-crystalline form[6,7,10,11], whereas epitaxial thin-film demonstrations remain exceedingly scarce. These limitations underscore the urgent need to identify environmentally benign, compositionally simple, and structurally stable compounds that inherently host dual topology and are compatible with scalable thin-film growth. Progress in this direction is essential for translating the coupled topology of real and reciprocal space into practical, device-relevant topological spintronic technologies.

The $Fe_5Si_3$ compound is emerging as a promising candidate for investigation as a topological ferromagnet, featuring a centrosymmetric hexagonal $D8_8$-type crystal structure with a space group of *P6₃/mmc*, indicating the presence of several interesting mirror and glide symmetries which may be responsible for non-trivial band topology[6]. While structurally analogous compounds, such as $Mn_5Si_3$[12], $Cr_5Si_3$[13], $Fe_5Sn_3$[11,14] and $Mn_5Ge_3$[6,15,16], have been extensively studied for their topological Hall effects and non-trivial electronic band structures, $Fe_5Si_3$ remains relatively unexplored. This is primarily due to the synthesis challenges posed by its non-equilibrium phase behaviour, particularly in epitaxial thin films and single-crystal forms. A recent study has highlighted the potential of $Fe_5Si_3$ in hosting spontaneous skyrmionic bubbles in nanorod configurations[17] across various temperature ranges. Another report predicts high spin angles, attributed to its significant uniaxial magneto-crystalline anisotropy[18]. Therefore, exploring $Fe_5Si_3$ as a topologically non-trivial and potentially safe ferromagnet could significantly advance the field of Topological Spintronics.

This study proposes $Fe_5Si_3$ as a potential magnetic Weyl nodal-line semimetal, based on comprehensive *ab initio* calculations. We conduct detailed experimental investigations of the Magneto-transport study on epitaxial thin films to elucidate the underlying magnetic textures. By integrating theoretical predictions with experimental observations, the study identifies a significant Berry curvature originating from Weyl nodal lines near the $E_F$, which accounts for the observed large intrinsic AHC of 504 S/cm at 200 K in a 50 nm epitaxial film. Furthermore, an AHA of approximately 5.5 % is recorded, placing $Fe_5Si_3$ among the magnetic Weyl semimetals known for their pronounced values of AHE. The system further hosts a sizable THE resistivity of 1.6 μΩ cm at 350 K, driven by its substantial uniaxial anisotropy. These findings highlight $Fe_5Si_3$ as a compelling platform for investigating topological spintronic phenomena and underscore its potential in future applications such as information storage and quantum computing technologies.

**Crystal Structure and First-principles Calculation**

$Fe_5Si_3$ crystallizes in a hexagonal $D8_8$-type structure (space group *P6₃/mmc*, No. 193) with lattice parameters $a = b = 6.759$ Å and $c = 4.72$ Å, exhibiting multiple mirror ($M_y$ and $M_x$), glide mirror ($G_x$, $G_y$ and $G_z$) and rotational ($C_{6z}$, $C_{3z}$, $S_{2z}$, $S_{2x}$ and $S_{2y}$) symmetries that are essential for the emergence of band crossings, Weyl points, and topological nodal lines. Figure.

1(a) shows the schematic crystal structure of $Fe_5Si_3$. The unit cell comprises of two inequivalent Fe sites and one Si site occupying distinct Wyckoff positions. The Fe (I) atoms reside at the *4d* Wyckoff position (1/3, 2/3, 0), forming chains along the *c*-axis with a 5-fold trigonal-bipyramidal $FeSi_5$ coordination. The Fe (II) atoms occupy the *6g* site at approximately ($x \approx$ 0.230, 0, 1/4), exhibiting an 8-fold coordination environment bonded to two neighbouring Fe and six Si atoms, thereby constituting the primary Fe sublattice. The Si atoms also occupy a *6g* site at (x $\approx$ 0.599, 0, 1/4), surrounded by nine Fe atoms forming a 9-fold coordination polyhedron. The distinct Wyckoff positions and coordination geometries of these atomic sites result in different local crystal fields and bonding characteristics, giving rise to *inequivalent* electronic and magnetic behaviours among the Fe sublattices.

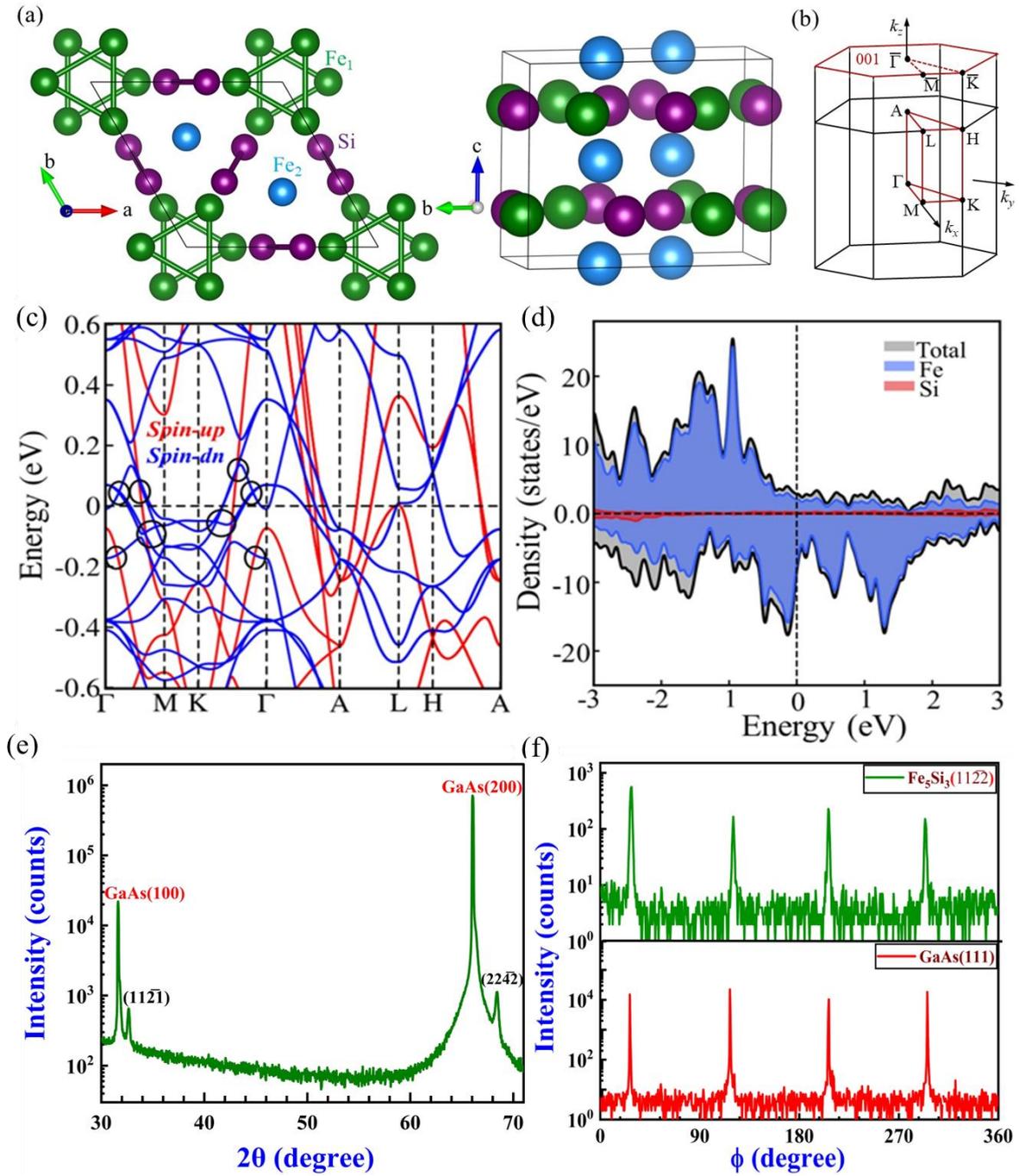

**Figure 1.** Crystal structure and structural characterization. (a) Schematic view of the crystal structure of $Fe_5Si_3$. (b) The irreducible Brillouin zone of the bulk, along with its projected (001) surface. (c) Spin-resolved electronic band structure without SOC. (d) Total and projected density of states of Fe and Si. (e) $2\theta$-$\omega$ scan of 50 nm films epitaxially grown on GaAs (100) substrate. (f) $\varphi$-scans of GaAs (111) peaks and $Fe_5Si_3$ ($11\bar{2}2$) peaks of the 50nm film.

The detailed electronic band structure of $Fe_5Si_3$ is obtained from first-principles DFT calculations. The spin-polarized band structure of $Fe_5Si_3$ is shown in Fig. 1(c). This compound exhibits metallic properties, featuring both spin-up and spin-down channels crossing the $E_F$. The projected density of states (PDOS) plot, given in Fig. 1(d), shows that the Fe-$d$ orbitals contribute predominantly, while the contribution from Si is negligible. We have identified several interesting band crossing points near the $\Gamma$-point, close to the Fermi level, along the high-symmetry $\Gamma$–$M$ and $K$–$\Gamma$ paths in the $k_z = 0$ plane for the spin-up channel and mixing of spin up and spin down channels which may lead to nodal line feature (see Fig. 1(c)). To emphasize the topological characteristics, the projected band structure is presented in Fig. S7(a). Notably, there is a mixing of orbital characters between the Fe-$d_{xz}$ and Fe-$d_{x^2-y^2}$ and Fe-$d_{z^2}$ and Fe-$d_{x^2-y^2}$ states near $E_F$ within the $k_z = 0$ plane, accompanied by opposite mirror eigenvalues of +1 and -1 at the crossing point. This hybridization leads to *band-inversion* with the inclusion of SOC, indicating the presence of *non-trivial* topological properties. For further confirmation of formation of nodal line feature without SOC, several evenly spaced paths between $M$ and $K$ points were also considered, as illustrated in Fig. 2(a). The analysis of the computed band dispersions (in reference to Fig. 2(b)) indicates that similar band crossings occur along the $\Gamma$–$M$/a/b/c/d/$K$ paths. This confirms the existence of a $\Gamma$-centred nodal line (NL) protected by $M_z$ and $G_z$ symmetries within the $k_z = 0$ plane, as schematically depicted in Fig. 1(b). Further confirmation of the presence of these nodal rings is provided by gap plane calculations on the (001) plane, as shown in Fig. 2(c). Apart from this, there are several nodal rings which are shown in the 2D Fermi surface plots (see Fig. S7). To confirm the non-trivial topological character, we have performed a Berry phase analysis and found a quantised value of +$\pi$, which unambiguously confirms that Fe$_5$Si$_3$ is a topologically non-trivial nodal-line semimetal. The 3D band structure of the nodal ring is illustrated in Fig. 2(d). In this case, magnetization along the [010] direction breaks the screw-axis rotation symmetries ($S_{2x}$ and $S_{2z}$) and the glide mirror symmetries ($G_x$ and $G_z$), while preserving the glide mirror symmetry $G_y$, and the screw-axis rotation symmetry $S_{2y}$. This symmetry breaking causes the gapped NL within $K_z = 0$ to evolve into hidden Weyl points, where six pairs of Weyl nodes are obtained, as presented in Table SII (see Supplementary Information). The Weyl points are marked in the BZ as shown in Fig. 4(h), and the corresponding band dispersions near these Weyl nodes are illustrated in Fig. S6. According to the Nielsen–Ninomiya no-go theorem[19], these twelve Weyl nodes collectively carry a net topological charge of zero. Each Weyl node acts as a magnetic monopole in momentum space, with a topological charge serving as either a source or a sink of the Berry curvature flux. This flux is determined by enclosing a surface around the Weyl node and takes a quantized value of +1 or −1, depending on the chirality of the Weyl fermion.

The Chern number associated with each Weyl node $W_i \pm$ (where $i$ = 1, 2, 3, 4, 5 and 6) is given by $C = \pm 1$.

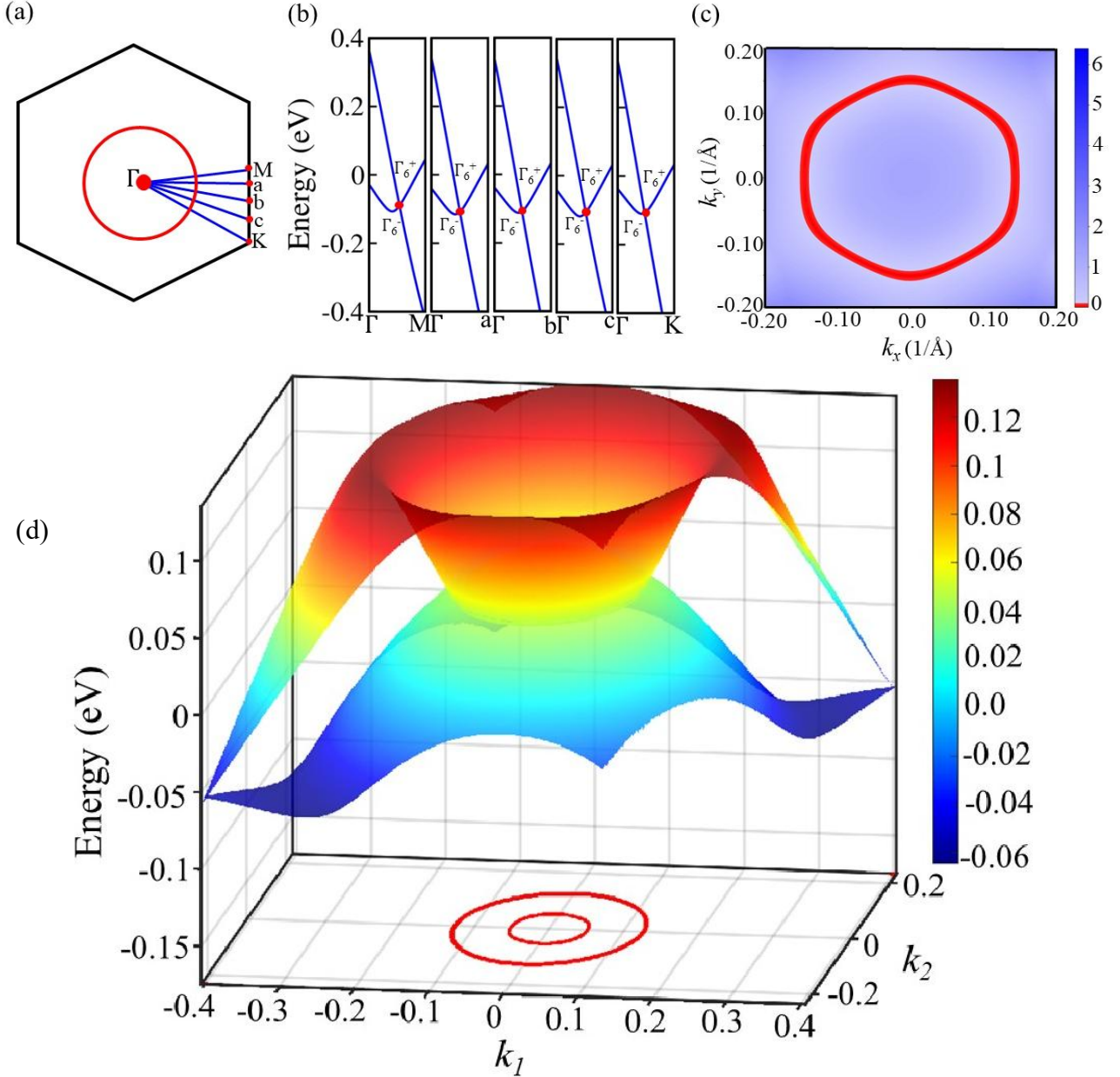

**Figure 2.** Topological Nodal line features. (a) Illustration of Nodal-line features, where *a-c* is equally spaced between *M* and *K*. (b) Electronic band structure along the *k*-paths indicated in (a). (c) The Nodal-ring dispersions along the (001) surface. (d) Three-dimensional band dispersions of the proposed Nodal-ring states.

## Structural characterization

To explore this Weyl nodal line material prediction, $Fe_5Si_3$ epitaxial thin films of nominal thickness 50 and 100 nm were deposited on GaAs (100) substrates by DC magnetron co-sputtering. High-purity (99.99 %) Fe and Si targets (2-inch diameter) were employed in a high-vacuum chamber evacuated to $< 5\times10^{-7}$ Torr via turbo and cryo-pumping. The GaAs substrates were sequentially ultrasonicated in acetone and isopropanol (each for 20 min), dried in $N_2$, and thermally descaled *in situ* at 800 °C for 30 min under high vacuum to remove native oxides. The films are deposited at a substrate temperature of 710 °C under a dynamic argon atmosphere maintained at $3\times10^{-3}$ Torr via a controlled gas flow rate of 64.9 sccm. The substrate rotation assembly was operated at 20 rpm to minimize thickness gradients and suppress growth-induced magnetic anisotropy. The films underwent *in situ* annealing at 710 °C for 90 minutes to enhance crystallinity. Additionally, a 3 nm aluminium capping layer was deposited at room temperature to prevent oxidation in the ambient environment.

The microstructural characterization of the $Fe_5Si_3$ thin film was carried out using X-ray diffraction (XRD) on a Malvern PANalytical Empyrean diffractometer equipped with a Cu-$K_\alpha$ radiation source ($\lambda$ = 1.5406 Å). To assess the crystalline quality and phase purity, a *2θ–ω* scan was performed on the $Fe_5Si_3$ film epitaxially grown on GaAs (100) substrate (Fig. 1(e)). The diffraction pattern exhibits distinct reflections corresponding to the ($11\bar{2}1$) family of planes, confirming the highly oriented epitaxial growth of the $Fe_5Si_3$ layer along the ($11\bar{2}1$) direction. The observed diffraction peaks are in excellent agreement with the JCPDS card number # 653593, corresponding to the hexagonal crystal structure of $Fe_5Si_3$. No additional peaks associated with any secondary, or impurity phases were detected, even under grazing incidence X-ray diffraction (GIXRD) measurements (Fig. S2(d) of SI), confirming the formation of a single-phase, well-crystallized film. Moreover, to further elucidate the epitaxial relationship between the $Fe_5Si_3$ film and the GaAs substrate, an asymmetric azimuthal ($\varphi$) scan was performed along the ($11\bar{2}2$) plane of $Fe_5Si_3$ by tilting the sample surface by 19.5° (Fig 1(f)). The $\varphi$-scan pattern exhibits four distinct peaks separated by 90°, which coincide with the 4-fold symmetry of the (111) reflections from the GaAs substrate, obtained by tilting the substrate surface by 54.7°. This correspondence confirms a well-defined in-plane epitaxial alignment between the $Fe_5Si_3$ ($11\bar{2}2$) film and the GaAs (100) substrate. Such a form of heteroepitaxy, involving an unusual crystallographic orientation of a hexagonal film on a cubic substrate, has also been reported in literature.[20,21]

To further confirm the elemental composition and chemical states of the $Fe_5Si_3$ thin film, Energy-Dispersive X-ray Spectroscopy (EDS) and high-resolution X-ray Photoelectron Spectroscopy (XPS) were performed. The EDS spectrum confirms Fe and Si as the primary constituents, with an atomic ratio close to the stoichiometric 5:3 (Figure S2(a)). In XPS spectra, the Fe 2p spectrum displays the $2p_{3/2}$ and $2p_{1/2}$ components at 706.9 and 719.9 eV, respectively (Fig. S2(c,e) of SI), characteristic of metallic Fe–Si bonding in $Fe_5Si_3$; weaker features at 710.4 and 712.7 eV are assigned to $Fe^{2+}$ and $Fe^{3+}$ species, in agreement with previous reports on $Fe_5Si_3$ nanostructures[22]. The coexistence of $Fe^{2+}$ and $Fe^{3+}$ valence states indicate the electron exchange between Fe and Si, thereby reflecting the charge balance and mixed covalent–metallic bonding character of the $Fe_5Si_3$ lattice. The presence of Si 2p doublet at 99.3 eV ($2p_{3/2}$) and 99.8 eV ($2p_{1/2}$) is consistent with Si–Fe bonding, while minor components at 102.4 and 105.6 eV are attributed to $SiO_2$ and sub-stoichiometric $SiO_x$ related to the native oxide[23,24]. Collectively, these data confirm the near-stoichiometric Fe:Si ratio and the characteristic covalent–metallic bonding framework of $Fe_5Si_3$.

**DC Magnetization behaviour**

Magnetization measurements were conducted on epitaxial $Fe_5Si_3$ thin films having 50 nm and 100 nm thicknesses using a Quantum Design MPMS3 superconducting quantum interference device (SQUID) magnetometer over a wide temperature range. The magnetic anisotropy was characterized through magnetization measurements performed in both in-plane (IP) and out-of-plane (OOP) configurations to evaluate the directional dependence of the magnetic properties. The temperature-dependent magnetization curves were obtained under zero-field-cooled (ZFC) and field-cooled (FC) protocols, with measurements conducted under an applied magnetic field of 500 Oe. The observed thermomagnetic irreversibility between FC and ZFC magnetization curves in the in-plane geometry provides evidence for establishing long-range magnetic ordering within the thin film samples (Fig. 3(a)). The temperature derivative of magnetization ($dM/dT$) analysis revealed a ferromagnetic to paramagnetic transition at approximately 370K, corresponding to the Curie temperature ($T_c$), which demonstrates consistency with previously reported values for both bulk and nanorod forms of $Fe_5Si_3$[17,25,26]. The OOP orientation exhibits a similar magnetic trend, consistent with the behaviour observed in the IP configuration (see Fig S3 of SI).

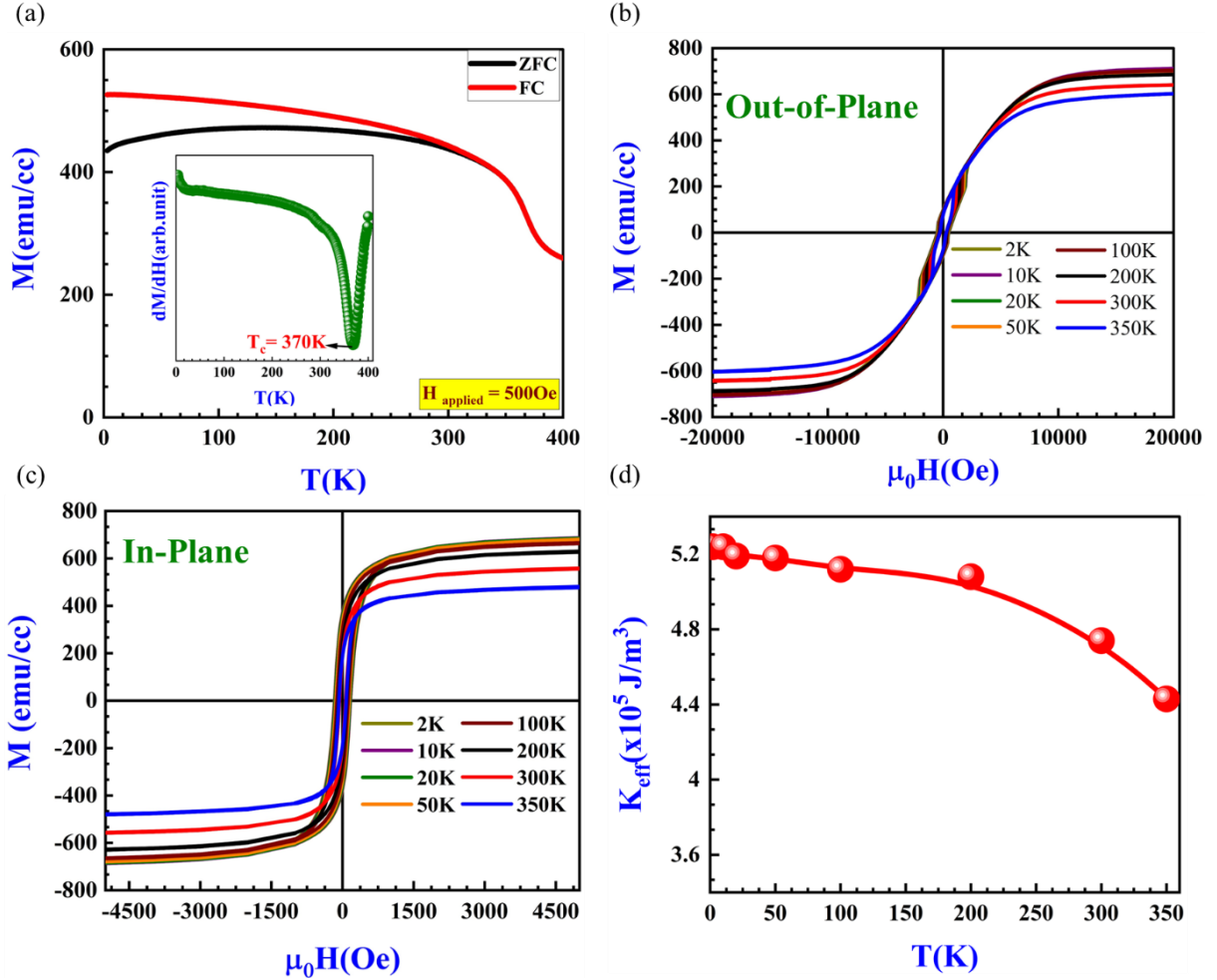

**Figure 3.** (a) Temperature-dependent magnetization under an in-plane magnetic field of 500 Oe. (b-c) Magnetization hysteresis loops of $Fe_5Si_3$ at different temperature under the IP and OOP direction. (d) Effective magnetic anisotropy constant of $Fe_5Si_3$ at different temperatures

Magnetic field-dependent magnetization (M–H) measurements were carried out at various temperatures for both IO and OOP orientations of $Fe_5Si_3$ thin films (Fig. 3(b,c)). At 2K, saturation magnetization ($M_S$) reached a value of approximately 700 emu/cm³ (7.4 $\mu_B$/ f.u.) under an applied magnetic field of 0.2 T along the IP direction. In contrast, saturation in the OOP direction required a significantly higher field of 1.3T, indicating strong magneto-crystalline anisotropy (Fig. S3(b) of SI). This anisotropic behaviour was consistently observed across the entire temperature range from 2 K to 350 K. The effective magnetic anisotropy constant ($K_{eff}$) was estimated to be 5.24 × 10⁵ J/m³ at 2 K (Fig. 3(d)), confirming the highly anisotropic nature of $Fe_5Si_3$. In addition to this, OOP M–H measurements reveal the presence of distinct multistep hysteresis loops around zero field, consistently observed across a wide range of temperatures and film thicknesses. Such behaviour suggests complex magnetization reversal mechanisms beyond conventional domain wall motion. Notably, recent study has reported the formation of spontaneous skyrmionic bubble states in $Fe_5Si_3$ nanorods[22], indicating the possible stabilization of similar topological spin textures in thin-film geometries. The discrete steps observed in the M–H loops are thus likely a consequence of abrupt polarity switching and collective dynamics of these topologically protected spin structures[27–29]. [22]

## Resistivity and Magneto-Transport Study

To investigate the magnetic Weyl characteristics and the Berry curvature–driven AHC, magneto-transport measurements were performed on epitaxial $Fe_5Si_3$ thin films with thicknesses of 50 and 100 nm. The temperature-dependent resistivity (Fig. 4(f)) exhibits a steady, monotonic increase with temperature, indicating robust metallic conduction, in agreement with prior reports[22,26] and supported by our DFT calculations. The residual resistivity ratio (RRR) of 2.8 attests to the high crystalline order of the films, reflecting minimal defect-mediated scattering. Furthermore, the 50-nm film exhibits a distinct negative and non-saturating longitudinal magnetoresistance (LMR) of ~2% at 200 K and 7 T (Fig. S4(a) of SI), *a response consistent with the chiral-anomaly–linked behaviour anticipated in a Weyl semimetal*[30–32].

The field-dependent anomalous Hall resistivity ($\rho_H^A$) was measured in epitaxial $Fe_5Si_3$ films over the 2–350 K temperature range under OOP field orientation up to 3 T (Fig. 4(d)). During the measurements, the electric current was applied along the in-plane direction, while the magnetic field was oriented perpendicular to the film plane. The details regarding data acquisition, symmetrization procedures, and the subtraction of the ordinary Hall effect (OHE) from the total Hall signal are provided in the Sec. 7 of SI. The $\rho_H^A$ values were obtained by extrapolating the high-field linear region of the Hall resistivity curves to zero magnetic field.

## **Theoretical Framework of the AHC calculation**

The schematic of the three-dimensional (3D) Berry curvature along the high-symmetry path, along with the two-dimensional (2D) Berry curvature arising from the broken nodal ring, is shown in Figs. 4(a, b). A sharp peak in the Berry curvature appears near the high-symmetry point $L$, along the band path, and a shallow valley is observed along the $\Gamma$–$A$ path, where gapped band crossings are present, as depicted in Fig. 4(a). The AHC of $Fe_5Si_3$ has been calculated from the Berry curvature using the following expression:

$$\sigma_{zx} = \frac{e^2}{\hbar} \int_{BZ} \frac{d^3k}{2\pi^3} \Omega_y^{zx}(\mathbf{k})$$

Here, $\sigma_{zx}$ denotes the anomalous Hall conductivity, $e$ is the elementary charge, $\hbar$ is the reduced Planck constant, and $\Omega_y^{zx}(k)$ represents the Berry curvature component integrated over the BZ. The calculated AHC at the $E_F$ is 333 S/cm, increasing to 805 S/cm at 0.05 eV, and reaching a maximum of 1400 S/cm at -0.77 eV. These relatively high values are comparable to those observed in other well-known topological systems[10,30,33–36].

Beyond the anomalous Hall response, the intrinsic spin Hall effect (SHE) provides an independent fingerprint of the SOC-driven electronic structure. Using a Wannier-interpolated tight-binding framework (see Supplementary Sect. S9), we find strongly anisotropic spin Hall conductivity (SHC) tensor components with maxima of 900 (ℏ/e) S/cm for $\sigma_{xy}^{spinz}$, 480 (ℏ/e) S/cm for $\sigma_{zx}^{spiny}$, and 1800 (ℏ/e) S/cm for $\sigma_{yz}^{spinx}$. These substantial values originate from SOC-induced anticrossings that generate concentrated spin Berry curvature hotspots throughout the Brillouin zone. Although a previous study attributed its observed large spin Hall angle (~0.15) to strong uniaxial magnetic anisotropy, the demonstration of high spin conversion efficiency is fully consistent with our prediction of sizable intrinsic SHC. Taken together, the pronounced AHC and intrinsic SHC indicate that $Fe_5Si_3$ could serve as an efficient ferromagnetic charge-to-spin converter, potentially usable for spin-pumping detection, inverse spin-Hall signal amplification, and integration into next-generation spintronic devices.

## **Temperature and Field Dependence of the Anomalous Hall Response**

As shown in Fig. 4(d), $\rho_H^A$ increases systematically with temperature from 2 K to 350 K. With increase in the strength of the applied magnetic field, $\rho_H^A$ initially rises in proportion to the magnetization and saturates around 1.2 T, consistent with the OOP magnetization behaviour. In conventional ferromagnets, the anomalous Hall resistivity typically follows the temperature dependence of magnetization and thus decreases with increasing temperature. However, in our epitaxial $Fe_5Si_3$ films, $\rho_H^A$ exhibits an *unconventional* enhancement with temperature, reaching a maximum value of ~7 μΩ cm at 350 K. Such an remarkable trend indicates the dominance of an intrinsic Berry curvature–driven mechanism, likely arising from temperature-induced shifts of the Fermi level across regions of large Berry curvature, as reported in several magnetic topological materials[34,36]. In addition to this, the AHC was calculated using $\sigma_H^A = \frac{\rho_H^A}{\rho_H^{A\,2} + \rho_{xx}^2}$ and the field and temperature dependent anomalous component $\sigma_H^A$ is plotted in Fig. 4(e). As shown in insert of Fig. 4(e), the $\sigma_H^A$ increases with temperature up to 200 K, attains a maximum of 504 S/cm, and then decreases, exhibiting a nonmonotonic behaviour that clearly defies linear scaling with $M$ and further supports an intrinsic, Berry-curvature-driven origin.

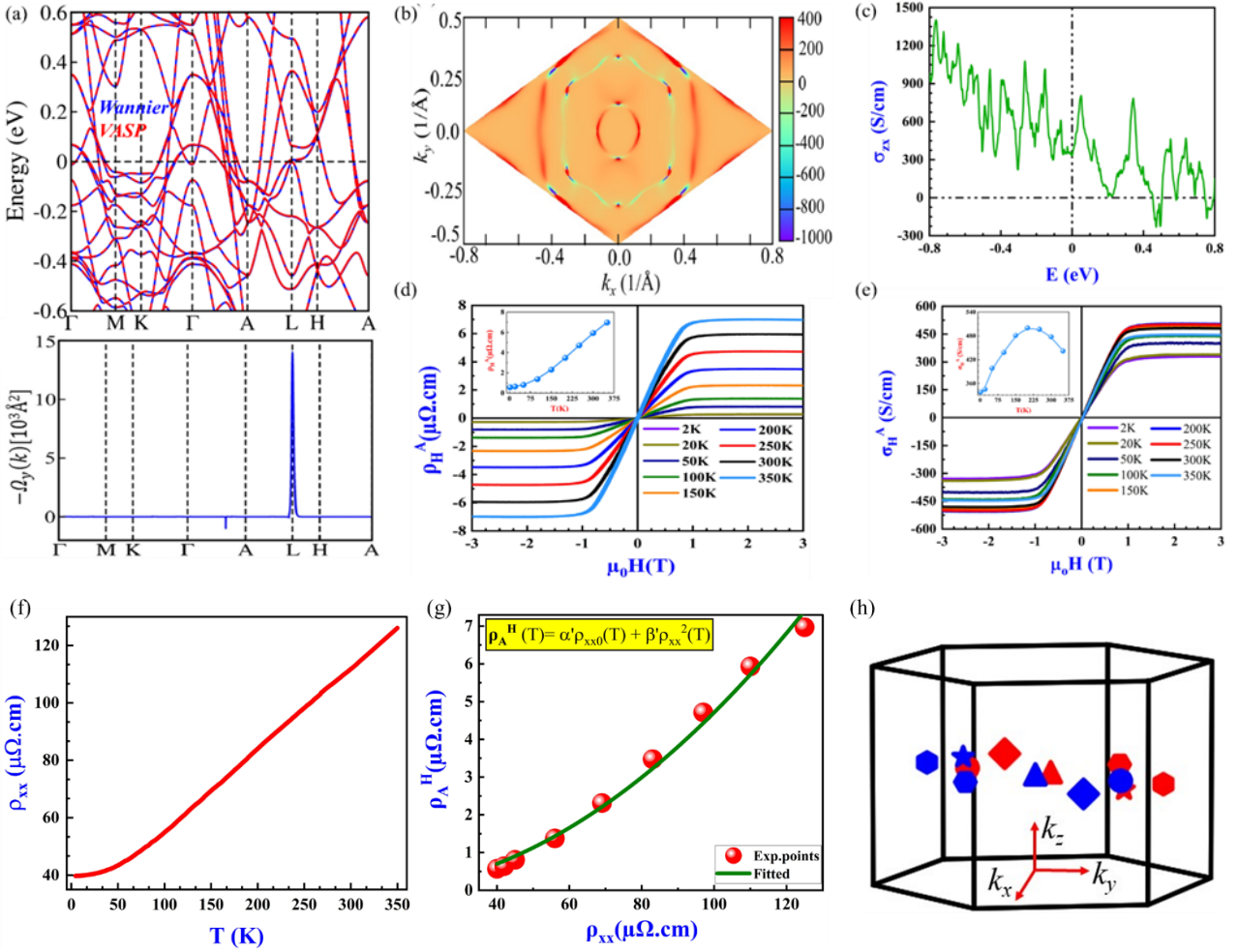

**Figure 4** (a) Berry curvature along the high-symmetry path in $Fe_5Si_3$. (b) Two-dimensional Berry curvature showing gapped nodal rings in $Fe_5Si_3$. (c) Calculated Anomalous Hall Conductivity $\sigma_{xz}$ with respect to Energy for $Fe_5Si_3$. (d & e) The Hall resistivity and Hall Conductivity at different temperatures. (f) Temperature-dependence of longitudinal resistivity of the 50nm thin $Fe_5Si_3$ film. (g) TYJ model analysis of 50nm thin $Fe_5Si_3$ film. (h) Positions of all 12 Weyl nodes in the bulk BZ, are shown for $Fe_5Si_3$. Red and blue indicate nodes with opposite chirality.

The AHC in ferromagnets originates from two distinct microscopic mechanisms: an intrinsic contribution governed by the Berry curvature of occupied electronic states in the BZ, and extrinsic contributions arising from spin–orbit–coupled scattering processes such as skew scattering and side-jump[37]. To determine the dominant mechanism present in these epitaxial $Fe_5Si_3$ films, the scaling relation between the anomalous Hall resistivity and the longitudinal resistivity is analyzed. A log–log plot of $\rho_H^A$ vs $\rho_{xx}$ at H = 2T (Fig. S5 (a) of SI) yields a slope of $b \approx 2.25$ ($\rho_H^A \propto \rho_{xx}^b$), signifying that the AHE is primarily governed by the quadratic (intrinsic/side-jump) mechanism. Further quantitative analysis using the Tian–Ye–Jin (TYJ) scaling model[38], $\rho_H^A = \alpha\rho_{xx0} + \beta\rho_{xx}^2$ yields a quadratic coefficient of β = (478 ± 17) S/cm, which accounts for nearly 95.4% of the maximum experimentally observed AHC (504 S/cm at 200 K) (fig. 4(g)). Experimentally, $\sigma_H^A$ increases from 340 S/cm at 2 K to a peak of 504 S/cm at 200 K and then decreases to 440 S/cm at 350 K. First-principles Berry-curvature calculations predict an intrinsic AHC of 333 S/cm (T = 0 K) at the Fermi level, in excellent agreement with the low-temperature value, confirming the intrinsic origin of the low-T Hall response. The observation that the fitted β slightly exceeds the DFT (T = 0 K) value yet closely matches the high-temperature magnitudes indicates that β acts as an effective quadratic coefficient encompassing temperature-dependent band structure and scattering effects. This behaviour is consistent with the calculated energy dispersion of $\sigma_H^A(E)$, at the Fermi level is 333 S/cm, increasing to 805 S/cm at 0.05eV, where non-trivial band crossing is present. Thermal broadening and subtle chemical-potential shifts at finite temperature can partially populate these high-Berry-curvature regions in momentum space, thereby enhancing the measured AHC, while side-jump scattering may further reinforce the quadratic contribution. Collectively, the experimental scaling, TYJ analysis, and first-principles results consistently establish that the anomalous Hall effect in epitaxial $Fe_5Si_3$ thin films is predominantly of intrinsic origin, driven by Berry curvature physics with only minor extrinsic scattering contributions.

The $Fe_5Si_3$ epitaxial film demonstrates another compelling topology driven Hall response: the anomalous Hall angle, AHA = $\frac{\sigma_H^A}{\sigma_{xx}}$, reaches a maximum of 5.5% at 350 K, while $\sigma_H^A$ rises to a maximum near 200 K. This behaviour can be understood from the distinct temperature dependences of the transverse and longitudinal channels. The $\sigma_H^A$ is governed by the Berry curvature integrated over the occupied bands, whose underlying topological structure is only weakly perturbed by thermal excitations below $T_c$. Therefore, $\sigma_H^A$ exhibits a high degree of temperature robustness. In contrast, the longitudinal charge conductivity ($\sigma_{xx}$) is strongly influenced by carrier scattering processes as with increasing temperature, enhanced electron–phonon interactions substantially reduce the carrier mobility and thus suppress $\sigma_{xx}$. This disparity between a topologically anchored transverse response and a scattering dominated longitudinal channel, naturally produces the observed enhancement of AHA over a wide temperature range below $T_c$ (Fig. 5(a)). In parallel, the anomalous Hall factor, $S_H = \frac{\sigma_H^A}{M}$, which estimates the relative magnitude of AHC with respect to magnetization, peaks at 0.072 V$^{-1}$ around 200 K, indicating the regime where Berry-curvature effects outpace the temperature-induced decline in magnetization (Fig. 5(b)). These interlinked temperature dependencies of $\sigma_H^A$, AHA and $S_H$ provide a coherent and quantitative transport signature of the Weyl nodal line feature in $Fe_5Si_3$.

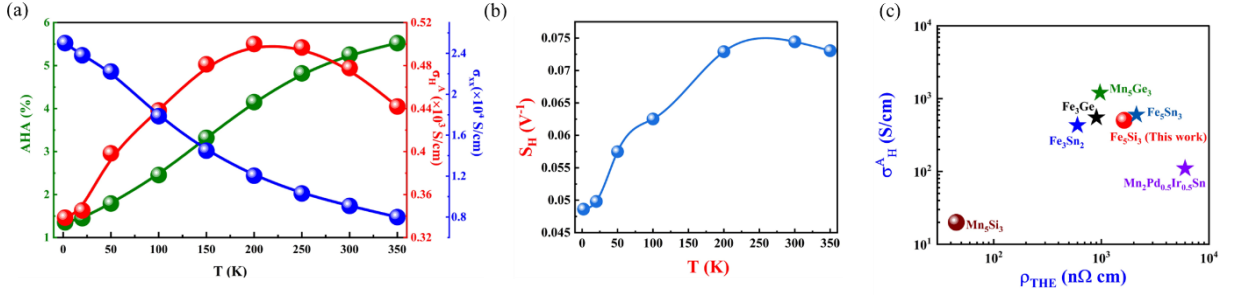

**Figure 5** (a) Temperature dependence of AHC, Longitudinal conductivity and AHA in the 50 nm thin $Fe_5Si_3$ film. (b) Temperature-dependent Anomalous Hall Factor $S_H$. (c) Comparison of AHC and THE resistivity of some of the previously reported well-known dual Topology materials. Here ★ symbol represents a single crystalline specimen, whereas ● represents a thin film specimen.

## Topological Hall effect

The THE arises from an emergent magnetic field generated by noncoplanar spin textures, such as skyrmions and multi-$q$ configurations, which impart a finite scalar spin chirality to the charge carriers[1,2]. Remarkably, these chiral textures can form even in centrosymmetric materials, where Dzyaloshinskii–Moriya interaction (DMI) is absent, through the combined influence of competing exchange interactions such as dipolar coupling, strong uniaxial anisotropy, and fluctuation-driven mechanisms[22,39–48]. Recent LTEM studies on $Fe_5Si_3$ nanorods have revealed spontaneous skyrmion-bubble states within a temperature range of 115–144 K, stabilized by the interplay between uniaxial anisotropy and dipolar interactions[17]. However, transitioning from a nanorod to a thin-film geometry substantially modifies the balance of magnetic interactions that govern texture stability[49,50]. Here, we investigate the topological Hall response of epitaxial $Fe_5Si_3$ thin films, providing evidence of topology-driven transport in this geometry, a feature that has been overlooked in earlier study.

Topological Hall resistivity ($\rho_{THE}$) was extracted over 2-350 K by decomposing the total Hall response into $\rho_H^A = R_0 H + R_s M + \rho_{THE}$, where subtraction of the ordinary ($R_0 H$) and anomalous ($R_s M$) components isolates the topological contribution (see Sec.7 of SI for details). Both 50 nm and 100 nm epitaxial $Fe_5Si_3$ films exhibit well-defined $\rho_{THE}$ hysteresis loops with finite remanence at zero field across the entire temperature range, demonstrating the spontaneous stabilization of noncoplanar spin textures in zero field, a behaviour also observed in several zero-field skyrmion-based systems[50–52]. The peak $\rho_{THE}$ increases steadily with temperature, reaching a remarkably high value of 1.63 μΩ·cm at 350 K for the 50 nm film, highlighting the notable thermal robustness of these chiral spin states. This behaviour contrasts sharply with earlier observations in $Fe_5Si_3$ nanorods, where such textures were restricted to a narrow window of temperature range i.e. 115–144 K. The substantially broadened stability range observed here underscores the critical influence of epitaxial thin-film geometry through symmetry breaking, modified uniaxial anisotropy and dipolar interactions, in reshaping the magnetic energy landscape and enabling the stabilization of topological spin textures far beyond the conventionally expected regime. These findings position epitaxial $Fe_5Si_3$ films as a promising platform, where geometry-engineered magnetic interactions yield high-temperature topological Hall signatures comparable to those of B20 compounds and other well-established skyrmion-hosting materials[7,8,50,52–54].

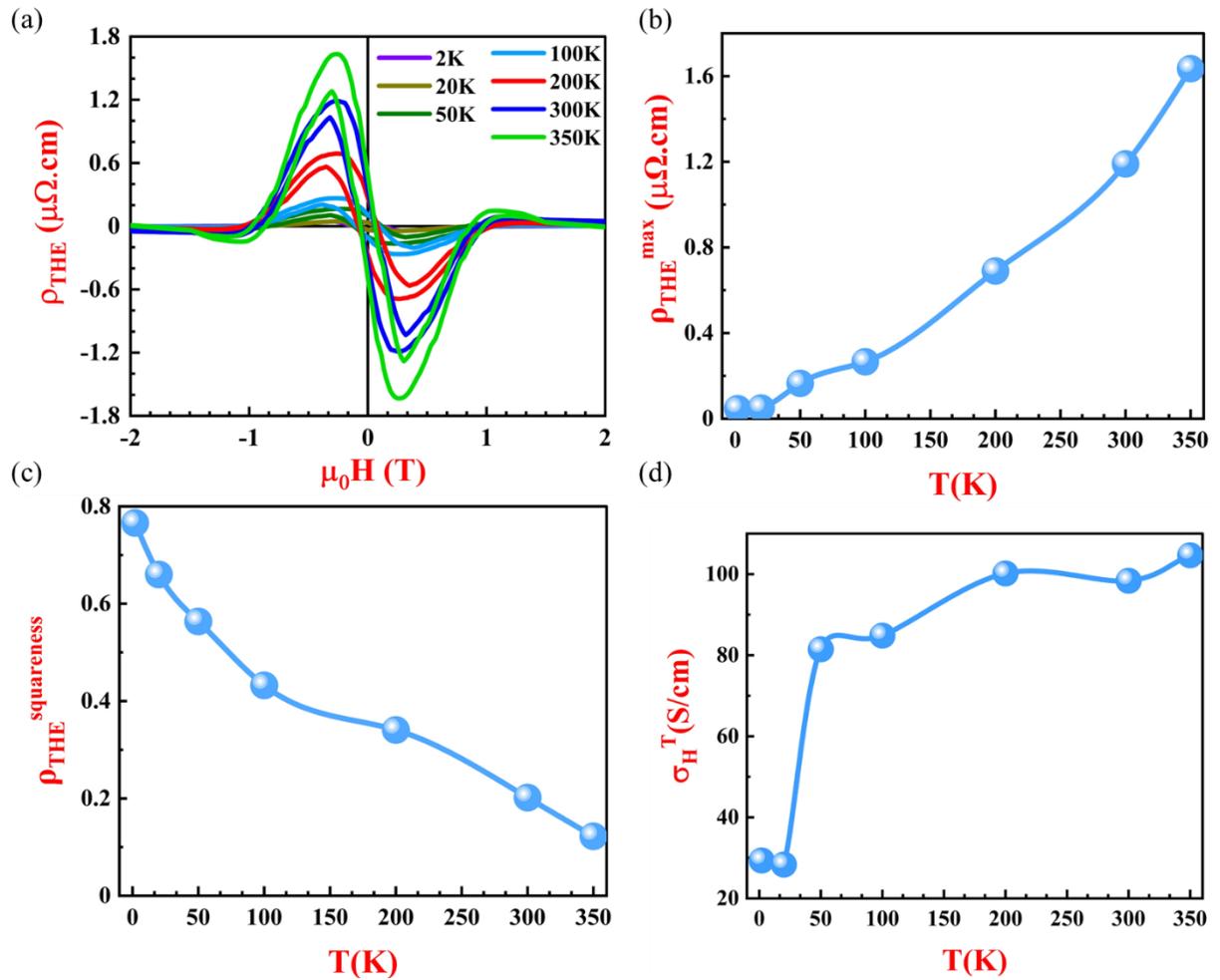

**Figure 6** (a) Topological Hall resistivity hysteresis loops of 50 nm thin $Fe_5Si_3$ film at different temperatures showing a substantial remanent signal at H = 0. (b) Temperature dependence of the maximum $\rho_{THE}$. (c) The squareness of the $\rho_{THE}$ of this 50nm thin $Fe_5Si_3$ film is indicative of the stability of the spin textures. (d) Temperature dependence of the calculated Topological Hall Conductivity $\sigma^T_H$ of 50 nm thin $Fe_5Si_3$ film.

We further quantify the intrinsic zero-field robustness of the chiral spin textures by evaluating the squareness ratio $S = \frac{\rho_{THE}(H=0)}{\rho_{THE}^{Max}}$ (Fig. 6(c)). The 50 nm film exhibits an exceptionally high squareness of 0.76 at 2 K, revealing that the majority of the topological Hall signal survives even in the complete absence of an external magnetic field. Although S gradually decreases with temperature (Fig. 6(c)), a pronounced squareness persists up to room temperature, demonstrating that the non-coplanar spin textures remain remarkably stable against thermal agitation. A similar temperature-dependent trend is observed in the 100 nm film, confirming that this robustness is intrinsic to the epitaxial $Fe_5Si_3$ system and not thickness-specific. The topological Hall conductivity (THC), obtained using $\sigma_{xy}^T \approx \frac{\rho^{THE}}{\rho_{xx}^2}$ valid for $\rho_{xy} \ll \rho_{xx}$, also increases systematically with temperature, attaining 100 S/cm at 300 K (Fig. 6(d)). This magnitude far surpasses the small THE signals characteristic of canonical MnSi and is comparable to the largest values reported in kagome and frustrated magnets[7]. The combination of large $\sigma_{xy}^T$, robust zero-field remanence and a broad thermal stability regime collectively establish epitaxial $Fe_5Si_3$ as a significant thin-film analogue of topological magnetic materials hosting dense, thermally resilient noncoplanar spin textures.

**Conclusion**

In summary, a comprehensive structural, magnetic, and magneto-transport investigation of epitaxial $Fe_5Si_3$ thin films, a newly predicted Weyl nodal-line material is presented. To the best of our knowledge, this work demonstrates for the first time the successful optimization of epitaxial $Fe_5Si_3$ films alongside *ab-initio* predictions of their magnetic Weyl semimetal character. The DFT calculations reveal a Weyl nodal line located near the Fermi level, accompanied by pronounced Berry curvature concentrated around high-symmetry points. Experimentally, a large anomalous Hall conductivity of ~504 S/cm at 200K and an anomalous Hall angle of ~5.5% at 350K are observed, which are in excellent agreement with theory, underscoring the dominant intrinsic Berry-curvature mechanism as supported by the TYJ scaling analysis. A clear negative longitudinal magnetoresistance signals the presence of the chiral anomaly, a hallmark of Weyl transport. Additionally, a sizable topological Hall resistivity of 1.63 μΩ cm at 350 K, along with the corresponding topological Hall conductivity of ~100 S/cm, and robust hysteretic THE loops, indicate the stabilization of chiral spin textures over a broad temperature range. Taken together, these results establish $Fe_5Si_3$ as a rare-earth-free, low-cost, high $T_C$ Weyl nodal-line system that uniquely hosts intertwined real-space and reciprocal-space topologies, offering a promising platform for advancing topological spintronics.

## Acknowledgments

S.P. acknowledges the University Grant Commission, Government of India, for financial assistance. We acknowledge the Institute's National Research Facility for EDAX measurements, the Central Research Facility for PPMS and XPS measurements, and the Physics Department Facilities for MPMS and XRD measurements. The author thanks Dr. Soumyarup Hait, Post-doctoral Research Fellow, University of Leeds for insightful discussions.